
%
%
%
%
%
%
%
\input uiucmac.tex

\PHYSREV
\sequentialequations
\tolerance 2000
\nopubblock
\titlepage
\title{\bf Renormalization
Group Theory for Global Asymptotic Analysis}
\author{Lin-Yuan Chen, Nigel Goldenfeld, and Y. Oono}
\address{Department of Physics, Materials Research Laboratory, and
Beckman
Institute, 1110 West Green Street, University of Illinois at
Urbana-Champaign,
Urbana, IL 61801-3080, U. S. A.}
\smallskip
\abstract
\noindent
We show with several examples that renormalization group (RG) theory
can
be used to understand singular and reductive perturbation methods in a
unified fashion.  Amplitude equations describing slow motion dynamics
in
nonequilibrium phenomena are RG equations.  The renormalized
perturbation approach may be simpler to use than other approaches,
because it does not require the use of asymptotic matching, and yields
practically superior approximations.

\medskip
\noindent
Pacs numbers: 47.20.Ky, 02.30.Mv, 64.60.Ak
\endpage

\noindent
The essence of the renormalization group (RG) is to extract those
structurally stable features of a system which are insensitive to
details\rlap.\REFS\nigelbook{N.\ D.\ Goldenfeld, {\sl Lectures on Phase
Transitions and the Renormalization Group} (Addison-Wesley, Reading,
Mass., 1992).}\REFSCON\kawasakifest{L.\ Y.\ Chen, N.\ Goldenfeld,
Y.\ Oono and G.\ Paquette, \journal Physica A &204&111(1994);
G.\ Paquette, L.\ Y.\ Chen, N.\ Goldenfeld, and Y.\ Oono, \journal
Phys.
Rev. Lett. &72&76(1994).}\refsend Thus, RG methods may be regarded as a
means of asymptotic analysis.  The usefulness of this point of view has
been amply demonstrated\Ref\rgpde{N.\ Goldenfeld, O.\ Martin and
Y.\ Oono, \journal J. Sci.  Comp.&4&355(89); N.\ Goldenfeld,
O.\ Martin,
Y.\ Oono and F.\ Liu, \journal Phys. Rev.  Lett.  &64&1361(90); J.
Bricmont and A.  Kupiainen, \journal Commun.  Math. Phys.
&150&193(92).}  by the relation between the RG and intermediate
asymptotics\rlap,\Ref\barenblatt{G.\ I.\ Barenblatt, {\sl Similarity,
Self-Similarity, and Intermediate Asymptotics} (Consultant Bureau, New
York, 1979).} which showed that the anomalous exponents appearing in
(\eg) the long-time behavior of certain hydrodynamic systems were
calculable using RG.

Many different techniques for asymptotic analysis have been developed
including the multiple scaling (MS) method (which actually subsumes all
the others), the boundary layer (BL) method, the WKB approximation, and
others\rlap.\Ref\bender{For example, C.\ M.\ Bender and S.\ A.\ Orszag,
{\sl Advanced Mathematical Methods for Scientists and Engineers}
(McGraw-Hill,1978); J.\ Kevorkian and J.\ D.\ Cole, {\sl Perturbation
Methods in Applied Mathematics} (Springer, New York, 1981).}  Reductive
perturbation methods\Ref\reductive{T.\ Taniuti and C.\ C.\ Wei,
\journal
J. Phys.  Soc. Jpn. &24&941(68); A.\ C.\ Newell and J.\ A.\ Whitehead,
\journal J. Fluid Mech. &38&279(69); Y.\ Kuramoto, {\sl Chemical
Oscillations, Waves, and Turbulence} (Springer, Berlin, 1984).} have
been used to extract the dynamics describing the global space-time
behavior of complicated systems near bifurcation
points\rlap.\Ref\crawford{See, \eg, J.D. Crawford, \journal Rev. Mod.
Phys. &63&991(91); M.C. Cross and P.C. Hohenberg, \journal Rev. Mod.
Phys. &65&851(93).}

At a purely technical level, the starting point for both perturbative
RG
methods and conventional asymptotic methods is the removal of
divergences from a perturbation series.  Given the above similarities,
a
natural question arises: what is the relation, if any, between
conventional asymptotic methods and the RG?

In this Letter, we demonstrate that many singular perturbation methods
may be understood as renormalized perturbation theory, and that
amplitude equations obtainable by the reductive perturbation methods
are
renormalization group equations\rlap.\Ref\nocon{We emphasize that our
RG
method has no connection with the so-called method of renormalization
or
uniformization\rlap,\refmark{\bender} which is a variant of the method
of stretched coordinates, and of limited use.}\ One of the advantages
of
the RG approach is that the starting point is simply a straightforward
naive perturbation expansion, for which very little {\it a priori}
knowledge is required; we will see that the RG approach seems to be
more
efficient and accurate, in practice, than standard methods in
extracting
global information from the perturbation expansion.

To illustrate the basic idea, let us consider a weakly nonlinear van
der Pol oscillator
$$
\frac{d^2 y}{d t^2}+y=\epsilon\Big\{\frac{d y}{d t}-\frac{1}{3}
\left(\frac{d y}{d t}\right)^3\Big\},\eqn\twelve
$$
which is usually solved by MS\rlap.\refmark{\bender} The method of
uniformization or `renormalization'  fails in this
case\rlap.\refmark{\bender}  A naive expansion $y=y_0+\epsilon
y_1+\epsilon^2 y_2+\cdots$ gives
$$
\eqalign{ y(t)=&R_0 \sin(t+\Theta_0)+\epsilon \Big\{A_1 R_0
\sin(t+\Theta_0) + \left(B_1-\frac{R_0^2}{96}\right)R_0
\cos(t+\Theta_0)\cr
 &+ \frac{R_0}{2}\left(1-\frac{R_0^2}{4}\right)(t-t_0)\sin(t+\Theta_0)
+\frac{R_0^3}{96}\cos3(t+\Theta_0)\Big\}+O[\epsilon^2],\cr}\eqn\thirteen
$$
where $R_0,\Theta_0,A_1,B_1$ are constants determined by the initial
conditions at arbitrary time $t=t_0$. This naive perturbation theory
breaks down for $\epsilon (t-t_0) > 1$ because of the secular terms.
The arbitrary time $t_0$ may be interpreted as the (logarithm of the)
ultraviolet cutoff in the usual field theory.\refmark{\rgpde} To
regularize the perturbation series, we introduce an arbitrary time
$\tau$, split $t-t_0$ as $t-\tau + \tau - t_0$, and absorb the terms
containing $\tau-t_0$ into the renormalized counterparts $R$ and
$\Theta$ of $R_0$ and $\Theta_0$, respectively.  This is allowed since
$R_0$ and $\Theta_0$ are no longer constants of motion in the presence
of the nonlinear perturbation. We introduce  a multiplicative
renormalization constant $Z_1=1+\sum_{1}^{\infty} a_n \epsilon^n$ and
an additive one $Z_2=\sum_{1}^{\infty} b_n \epsilon^n$ such that
$R_0(t_0)=Z_1(t_0,\tau) R(\tau)$ and $\Theta_0(t_0)=\Theta(\tau)
+Z_2(t_0,\tau)$.  The coefficients $a_n$ and $b_n$ ($n\ge 1$) are
chosen order by order in $\epsilon$ to eliminate the terms containing
$\tau-t_0$ as in the standard
RG\rlap.\Ref\gellman{E.\ C.\ G.\ Stueckelberg and A.\ Petermann,
\journal Helv. Phys. Acta &26&499(53); M. Gell-Mann and F. E. Low,
\journal Phys. Rev. &95&1300(54); J. Zinn-Justin, {\sl Quantum Field
Theory and Critical Phenomena} (Clarendon Press, Oxford, 1989).} The
choice $a_1=-(1/2)(1-R^2/4)(\tau-t_0) -A_1$, $b_1=-B_1$ removes the
secular terms to order $\epsilon$, and we obtain the following
renormalized perturbation result\Ref\rgconst{$Z_1$ may depend upon $R$,
because $R$ is dimensionless.  This is analogous to the renormalization
of a dimensionless coupling constant in field theory.}
$$
\eqalign{ y(t)=&\Big\{R+\epsilon
(R/2)(1-R^2/4)(t-\tau)\Big\}\sin(t+\Theta)\cr
&-\epsilon(1/96)R^3 \cos(t+\Theta) +\epsilon (R^3/96)
\cos3(t+\Theta)+O[\epsilon^2],\cr}\eqn\fourteen
$$
where $R,\Theta$ are now functions of $\tau$.   Since $\tau$ does not
appear in the original problem, the solution should not depend on
$\tau$.
Therefore, $(\partial y/\partial \tau)_t= 0$.  This is the RG equation,
which in this case consists of two independent equations
$$
d R/d \tau=\epsilon (R/2)\left(1-R^2/4\right)+O[\epsilon^2] ,\quad d
\Theta/d t=O[\epsilon^2].\eqn\fifteen
$$
Setting $\tau = t$ in \fourteen\ eliminates the secular term, giving
$$
y(t)=R(t)\sin(t)+(\epsilon/96) R(t)^3 \cos(3t)+O[\epsilon^2],
\eqn\seventeen
$$
where $R(t)$ and $\Theta(t)$ are obtained from \fifteen\ with $t =
\tau$: $R(t) = R(0)[e^{-\epsilon t} + R(0)^2(1-e^{-\epsilon
t})/4]^{-1/2}$ and $\Theta (t) = \Theta (0)$, which we take to be
zero.
The final result approaches a limit circle of radius tending towards 2
as $t\rightarrow \infty$. The second order RG calculation shows that
our
assumption of perturbative renormalizability is consistent and no
ambiguity arises.   The multiple time scales used in MS for slow
variables $T_1=\epsilon t, T_2=\epsilon^2 t, \cdots$ appear
naturally\rlap.\Ref\woodruff{A related result has been obtained
independently by S.L. Woodruff, \journal Studies in Applied Mathematics
&90&225(93).}

The above example illustrates two important points: (1) the results of
the MS method can be obtained from renormalized perturbation theory,
and (2) the RG equation describes the long time scale motion of the
amplitude and the phase.  In the following we wish to demonstrate (1)
and (2) more generally.

Another important class of singular problems is that for which the
highest order derivative of the equation is multiplied by a small
parameter $\epsilon$, \eg\ WKB and BL problems.  For linear cases, it
is known\refmark{\bender} that both problems can be treated in a
unified fashion. A typical problem is of the form
$$
\epsilon^2 \frac{d^2 y}{d x^2}+ a(x)\frac{d y}{d x}-b(x) y=0, \quad
0\le x \le 1, \quad \epsilon \rightarrow 0_{+}, \eqn\lone
$$
where  $a$ and $b$ are continuous functions, and we have chosen
$a(x)>0$ so that the boundary layer is at $x = 0$. This can be
transformed into a form suitable for WKB analysis:
$$
\epsilon^2 \frac{d^2 u}{d x^2}= Q(x) u(x), \eqn\lthree
$$
with $Q(x)\equiv a^2(x)/4\epsilon^2+a'(x)/2+b(x)$ and
$$
y(x)=\exp \left[-\frac{1}{2\epsilon^2}\int^{x} a(x')dx'\right] u(x).
\eqn\ltwo
$$
It is convenient to
introduce a new variable $t$
such that $dt=\sqrt{Q} dx/\epsilon$, and
\lthree\ becomes
$$
\frac{d^2 u}{d t^2} - u = -\frac{1}{2}Q^{-3/2}(x(t)) Q'(x(t))
\frac{d u}{d t},
\eqn\lfive
$$
where $Q'(x(t))$ is assumed to be a function of order unity varying
slowly on the time scale $t$, and $Q(x)\neq 0$ holds for $0\le x \le 1$
for simplicity.  Naively expanding $u$ as $u(t)=u_0(t)+\epsilon
u_1(t)+\cdots$, we find the bare perturbation result
$$
\eqalign{ u(t)=& e^{t}\Big\{A_0 + \epsilon A_0
\int_{t_0}^{t}S\left(x(s)\right) ds-\epsilon A_0
e^{-2t}\int_{t_0}^{t}S\left(x(s)\right) e^{2s}ds\Big\}\cr &
+e^{-t}\Big\{B_0 + \epsilon B_0 \int_{t_0}^{t}S\left(x(s)\right)ds
-\epsilon B_0 e^{2t}\int_{t_0}^{t}S\left(x(s)\right) e^{-2s}ds\Big\}
+O[\epsilon^2],\cr}\eqn\lsix
$$
where $S(x)\equiv -Q^{-3/2}Q'(x)/4$, and $A_0, B_0$ are constants
dependent on the initial conditions at $t=t_0$.  The terms in the curly
brackets are the secular terms divergent in the limit $t-t_0
\rightarrow \infty$.  These terms are renormalized away with the aid of
the multiplicative renormalization $A_0=Z_1 A(\tau)$ and $B_0=Z_2
B(\tau)$, where $A, B$ are the renormalized counterparts of $A_0, B_0$,
respectively.  Here $Z_1= 1 +\epsilon \int_{\tau}^{t_0}
S\left(x(s)\right)ds +\cdots = Z_2$, with $\tau$ being some arbitrary
time, as in the example of the van der Pol oscillator.  The
renormalized perturbation result is
$$
u(t)=e^{t}\Big\{A(\tau) + \epsilon A(\tau) \int_{\tau}^{t}S ds\Big\}
+e^{-t}\Big\{B(\tau) + \epsilon B(\tau) \int_{\tau}^{t}S
ds\Big\}+O[\epsilon],\eqn\lseven
$$
where $O[\epsilon]$ refers to all regular terms of order $\epsilon$
which remain finite even as $t-t_0 \rightarrow \infty$.  The RG
equation ${\pd u}/{\pd \tau}\equiv 0$ gives
$$
\frac{d C}{d \tau}+\epsilon \frac{1}{4}Q^{-3/2}Q'(x(\tau))
C=O[\epsilon^2], \eqn\leight
$$
where $C=A$ or $B$. Again, \leight\ corresponds to the
amplitude equation with $\tau=t$, which gives
$C(x)\sim Q^{-1/4}(x)$.  This is just the adiabatic invariant
$A(x)Q^{1/4}(x)=A(0)Q^{1/4}(0)=$ constant.  Therefore, the
`physical-optics' approximation for \lthree\ reads
$$
u(x)\sim Q^{-1/4}(x) \left\{C_1 \exp\left[\frac{1}{\epsilon}\int^{x}
dx'\sqrt{Q(x')}\right] +C_2 \exp\left[-\frac{1}{\epsilon}\int^{x}
dx'\sqrt{Q(x')}\right] \right\},\eqn\leleven
$$
where $C_1$ and $C_2$ are integration constants.
The asymptotic result $y(x)$ uniformly valid for $\epsilon x < 1$ for
the general linear boundary-layer problem \lone\ is given by
\ltwo\ with
an appropriate asymptotic expansion formula for $Q(x)$.  Notice that
the
above RG approach gives uniformly reliable results from the
``inner expansion'' alone without any (intermediate asymptotic)
matching.

We have seen that the RG equation becomes the equation of motion for
the slow behavior of the system.  To see how RG reproduces reductive
perturbation results, let us consider the following equation (this type
covers most examples so far studied in the literature)
$$
[{\cal L}_1(\partial_t) + {\cal L}_2(\nabla)] u = \epsilon Q[u],
\eqn\ichi
$$
where ${\cal L}_1$ and ${\cal L}_2 $ are constant coefficient linear
differential operators, $Q$ is a possibly nonlinear operator, and
$\epsilon$ is a small parameter.  We assume, for simplicity, spatial
isotropy.  Suppose that the operators have the following structures
${\cal L}_1 = \prod_{\omega}(\partial_t + i\omega)^{m(\omega)}$, ${\cal
L}_2 = \prod_{\mu}(\nabla - i \mu)^{n(\mu)}$, and $u_{0} =
\sum_{\omega,\mu} a_{\omega,\mu} e^{i(\mu\cdot x - \omega t)}$ is a
solution to \ichi\ with $\epsilon =0$\rlap.\Ref\exclu{Those $\mu$ and
$\omega$ which are not real do not give a bounded, globally meaningful
zeroth order solution, so they must be excluded from the expression for
$u_0$.}\
The order
$\epsilon $ correction, $u_1$, in the naive perturbation obeys $ ({\cal
L}_1 + {\cal L}_2) u_1= Q[u_0]. $ We assume without any loss
of generality that $Q[u_0]$ can be expanded as
$$
Q[u_0] = \sum_{\omega,\mu} Q_{\omega,\mu}[\{a\}] e^{i(\mu\cdot x -
\omega t)} + R,\eqn\qqone
$$
where $R$ is the remainder such that ${\cal L}_1 R \neq 0$, ${\cal L}_2
R \neq 0$, and $Q_{\omega,\mu}$ are coefficients dependent on the set
of $a_{\omega,\mu}$'s collectively denoted by $\{a\}$.  The general
form of the singular (secular) part $[u_1]_s$ of the general solution
is
$$\eqalign{
[u_1]_s =& \sum_{\omega,\mu} Q_{\omega,\mu}[\{a\}] \left\{ \lambda
\ell_{1}(\omega)^{-1}\frac{t^{m(\omega)}-t_0^{m(\omega)} }{m(\omega)!}
\right. \cr  &+ \left. (1-\lambda) \ell_{2}(\mu)^{-1}\frac{
|x|^{n(\mu)} - |x_0|^{n(\mu)} }{n(\mu)!} + P_{\omega,\mu} \right\}
e^{i(\mu\cdot x - \omega t)}, \cr}\eqn\singpart
$$
where $\lambda$ is an arbitrary numerical constant, not equal to $0$ or
$1$, $P_{\omega,\mu}$ is a polynomial of $t$ and $|x|^2$ of lower order
than $m(\omega)$ and $n(\mu)/2$, respectively (whose explicit form is
not required), $\ell_1(\omega) = \prod_{\omega' \neq \omega} (i \omega
- i\omega')^{m(\omega')} $ and $\ell_2(\mu) = \prod_{\mu' \neq \mu} (i
\mu' - i \mu)^{n(\mu')}$.  Renormalization of the secular terms
divergent in the global space-time limit can be done following the
procedures given above, and is tantamount to replacing in
\singpart\ $t^n-t_0^n$ with $t^n - \tau^n$, $|x|^n $ with $|x|^n -
|r|^n$  and the `bare' coefficients $\{a\}$ with their renormalized
counterparts $\{A\}$, regarded as functions of $\tau$ and $r$.  The
renormalization group equation can be obtained from the condition that
$u$ is independent of the parameters $\tau$ and $r$ introduced in the
renormalization process.  The term $P$ is dependent upon \eg, initial
conditions.  Thus, to obtain a universal result, we differentiate $u$
sufficiently many times with respect to $\tau$ and $r$ to eliminate
$P$.  Further eliminating $\lambda$, we find the following mode-coupled
amplitude equation:
$$
\ell_1(\omega)\frac{\partial^{m(\omega)} A_{\omega,\mu}}{\partial
t^{m(\omega)}}+\ell_2(\mu)\Delta^{n(\mu)/2}
A_{\omega,\mu} =
\epsilon Q_{\omega,\mu}(\{A\}).\eqn\modecoupl
$$
Here we have used the isotropy to introduce the Laplacian $\Delta$.
This is the renormalization group equation independent of the
arbitrariness due to initial conditions, solution methods, \etc

As an example, consider the Swift-Hohenberg
equation\rlap:\Ref\swift{J.\ Swift and P.\ C.\ Hohenberg, \journal
Phys. Rev. A &15&319(77).}
$$
\pd u/\pd t=\epsilon u -\left(1+\nabla^2 \right)^2 u -u^3,\eqn\nineteen
$$
where $\epsilon$ is a control parameter.  Here ${\cal L}_1 =
\partial_t$, ${\cal L}_2 = (1+\nabla^2)^2$, $\epsilon Q[u] = \epsilon u
-u^3$ and $u_0 = a e^{ix}+a^*e^{-ix}$.  Thus $\ell_1 = 1$, $\ell_2(\pm
1) = -4$, $\epsilon Q_{0,1}(A,A^*) = \epsilon A -3|A|^2A$ and $\epsilon
Q_{0,-1}(A,A^*) = \epsilon A^* -3|A|^2A^*$.  That is,
$$
\frac{\pd A}{\pd t}=4\frac{\pd^2 A}{\pd x^2}+ (\epsilon A-3|A|^2
A).\eqn\twentyfive
$$
Scaling out $\epsilon$ identifies the slow time and spatial variables
as $T=\epsilon t, X=x/2 \sqrt{\epsilon}$.  The scaled result is the
well-known time-dependent Ginzburg-Landau amplitude equation. Because
the secular terms $\epsilon t$ and $\epsilon x^2$ have been removed
from \singpart\, the outcome \twentyfive\ should be uniformly valid up
to time scale $1/\epsilon$ and spatial scale $1/\sqrt{\epsilon}$ for
$\epsilon \ll 1$. In our approach, spatial and time coordinates are
treated on an equal footing, in contrast to the standard reductive
perturbation method\rlap.\refmark{\reductive}

To demonstrate that there are not only conceptual but also technical
advantages to the RG approach, we conclude with a problem involving the
so-called `switchback': conventionally, only through subtle analysis in
the course of actually solving the problem is it possible to realize
the need for (\eg) unexpected order terms to make asymptotic matching
consistent.  An example is a caricature of the Stokes-Oseen
singular boundary layer problem, which describes the low Reynolds
number
viscous flow past a sphere of unit radius.  After scaling the radial
coordinate $r$ to $x = \epsilon r$, where $\epsilon$ is the Reynolds
number squared, the equation is\Ref\hinch{E.\ J.\ Hinch, {\sl
Perturbation Methods} (Cambridge University Press, Cambridge, 1991),
section 5.2.}
$$
\frac{d^2 u}{d x^2}+\frac{2}{x}\frac{d u}{d x}+ u \frac{d u}{d x}
=0, \qquad u(x=\epsilon)=0, \,\, u(x=\infty)=1. \eqn\stokes
$$
We regard \stokes\ as an initial-value problem, given an initial
condition $u(x_0)=A_0$ at some arbitrary point $x=x_0$, where $A_0$ is
a finite constant.  Assuming a naive expansion
$u(x;\epsilon)=u_0(x)+\lambda_{1}(\epsilon)u_1(x)
+\lambda_{2}(\epsilon)u_2(x)+\cdots$, where $\lambda_{i}(\epsilon),
i=1,2,\cdots$ will be determined in a self-consistent way, we obtain
$u_0(x)=A_0$. Solving to $O[\lambda_1(\epsilon)]$ gives
$$
u(x)=A_0 + \lambda_1(\epsilon) A_1 A_0\big \{e_{2}(A_0 x_0)-e_{2}(A_0
x)\big\}+ O[\lambda_1(\epsilon)^2, \lambda_2(\epsilon)],\eqn\third
$$
where $A_1(x_0)$ is some constant of integration, and we define
$e_{2}(t)\equiv \int_{t}^{\infty}d\rho
\rho^{-2}e^{-\rho} \sim 1/t+\log{t}+(\gamma-1) -t/2+O(t^2)$ as $t
\rightarrow 0$ with Euler's constant $\gamma\simeq 0.577$.  The naive
perturbation result \third\ breaks down when both $x_0$ is small and
$x-x_0$ is large.  To cure this we introduce  the renormalization
constant $Z= 1 - \lambda(\epsilon) A_1 \{e_{2}(A\mu)-e_{2}(A x_0)\}$
such that $A(\mu) = Z A_0$, giving the renormalized perturbation series
$$
u(x)=A(\mu) + \lambda_1(\epsilon) A_1 A\big \{e_{2}(A \mu)-e_{2}(A
x)\big\}+ O[\lambda_1(\epsilon)^2, \lambda_2(\epsilon)].\eqn\fourth
$$
The RG equation ${\pd u}/{\pd \mu}=0$ yields, after setting $\mu = x$,
$$\frac{d A(x)}{d x}=\lambda_1(\epsilon) A_1 \frac{e^{-A(x) x}}{x^{2}}
+ O[\lambda_1(\epsilon)^2, \lambda_2(\epsilon)],\eqn\fifth
$$
and we find the final uniformly valid asymptotic result $u(x)=A(x)$.

Equation \fifth\ can be solved iteratively along with
the required boundary conditions $A(\infty)=1$ and $A(\epsilon)=0$ as
$$
A(x)=1-\lambda_1(\epsilon) A_1 \int_{x}^{\infty} d \rho
\rho^{-2}e^{-\rho} + O[\lambda_1(\epsilon)^2,
\lambda_2(\epsilon)].\eqn\sixth
$$
The condition $A(\epsilon)=0$ gives $\lambda_1(\epsilon) A_1
e_{2}(\epsilon) =1$. Therefore, the expansion coefficient $\lambda_1$
can be explicitly chosen as $\lambda_1(\epsilon)A_1=1/e_{2}(\epsilon)$,
with the asymptotic expansion, in the limit $\epsilon\rightarrow
0_{+}$,
$A_1\lambda_1(\epsilon)\sim \epsilon (1-\epsilon \log
{\epsilon}-(\gamma-1)\epsilon+\cdots)$.  In addition, we require that
${\lambda_1}^2/\lambda_2 =O(1)$, so that the equation for $u_2$ yields
new information.  The resulting asymptotic solution is correct to
$O[\epsilon\, \log \epsilon]$ and agrees with that obtained by
asymptotic matching.   Note that in our method, the $\epsilon
\log{\epsilon}$ term appears naturally from the asymptotic expansion of
$e_2(\epsilon)$, whereas some artistry is required to obtain this term
conventionally.    The result to $O[\epsilon]$ given by asymptotic
matching\refmark{\hinch} is obtained from the renormalized perturbation
expansion to $O[\lambda_2]$.  The asymptotic expansion to $O[\epsilon]$
is not uniformly valid in $r$, and a much better approximation, in
practice, is our full result $1-e_2(\epsilon r)/e_2(\epsilon)$ to order
$\lambda_1$.
This can be seen clearly in figure \FIG\figi{Comparison between the
numerical solution of eq. \stokes\ for several values of $\epsilon$,
the
first order RG result $1-e_2(\epsilon r)/e_2(\epsilon)$, and two
matched
asymptotic expansions (one at fixed $r$, the other at fixed $\rho\equiv
r\epsilon$), as derived in ref. \hinch.}\figi.

In summary, we have demonstrated that various singular perturbation
methods and reductive perturbation methods may be understood in a
unified fashion from the renormalization group point of view, with some
attendant technical advantages.

We are grateful to Paul Newton for valuable discussions.  LYC
acknowledges the support of grant NSF-DMR-89-20538, administered
through
the University of Illinois Materials Research Laboratory. NG and YO
gratefully acknowledge partial support by National Science Foundation
Grant NSF-DMR-93-14938.

\refout


\figout


\end